\documentclass[aps, twocolumn, superscriptaddress]{revtex4}%

\usepackage{epsfig,dsfont,amssymb,amsmath,amsthm,amsfonts,amsbsy,mathrsfs}
\usepackage{graphicx, txfonts}
\usepackage{bm}

\begin{document}

\title{Continuous-variable entanglement generation using a hybrid $\mathcal{PT}$-symmetric system}
\author{Saeid Vashahri-Ghamsari}
\affiliation{
Department of Physics,
University of Arkansas,
Fayetteveille, AR 72701, USA
}
\author{Bing He}
\email{binghe@uark.edu}
\affiliation{
Department of Physics,
University of Arkansas,
Fayetteveille, AR 72701, USA
}
\thanks{binghe@uark.edu}
\author{Min Xiao}
\email{mxiao@uark.edu}
\affiliation{
Department of Physics,
University of Arkansas,
Fayetteveille, AR 72701, USA
}
\affiliation{National Laboratory of Solid State Microstructures and School of Physics, Nanjing University, Nanjing 210093, 
China}

\begin{abstract}
\noindent We study a proposal of generating macroscopic continuous-variable entanglement with two coupled waveguides respectively carrying optical 
damping and optical gain. Moreover, a squeezing element is added into one or both waveguides. We show that quantum noise effect existing in the 
process is essential to the degree of the generated entanglement. It will totally eliminate the entanglement in the setup of adding the squeezing 
element into the waveguide filled with optical damping material, but will not completely damp the entanglement to zero in 
the other configuration of 
having the squeezing element in the gain medium or in both gain and damping medium. The degree of the generated continuous-variable entanglement is irrelevant to the intensities of the input light in coherent states.
Moreover, the relations between the entanglement and system parameters are illustrated in terms of the 
dynamical evolutions of 
the created continuous-variable entanglement.
\end{abstract}

\maketitle

\section{Introduction}

\noindent Entanglement, a purely quantum mechanical feature that distinguishes
quantum systems from their classical counterparts,
has various important applications in quantum computing and
quantum information science \cite{1}. 
Creating the necessary entangled quantum states in feasible ways is always a prerequisite in these applications.

Nowadays, there are two main approaches to the quantum information processing: (1) the ``digital" approach, 
in which the information is encoded in quantum systems with discrete degrees of freedom (qubits or qudits) such as two polarization states of a single
photon, spin $1/2$ electrons, and two lowest energy levels of
quantum dots; (2) the 
``analog" approach, in which the quantum correlations are encoded in continuous-variable degrees of freedom (CVs) 
such as the quadrature amplitude of quantized harmonic oscillator, especially those of the electromagnetic field, as well as Josephson junction and Bose-Einstein condensate \cite{3,4}.

Light is a good carrier of quantum information as it interacts weakly with the environment. Quantum information encoded in CV states of light enjoys numerous advantages of preparing, manipulating, and measuring, as compared with the photonic qubits \cite{3}. CV quantum states are often in the form of Gaussian states, the manipulation of which is within the reach of current experimental technology. In addition, quantitative description of all properties of the Gaussian states is possible 
\cite{ade}. These benefits motivate one to explore the ways of generating entangled Gaussian states.

Entanglement involving light fields with high intensity is an example of the so-called macroscopic entanglement. 
In addition to possible applications \cite{3,4}, generating light fields of macroscopic entanglement or 
macroscopic superposition is importantly meaningful in fundamental physics; see, e.g. the recent experimental \cite{ex0, ex1} and theoretical studies \cite{a1, Raeisi, a2, a3}. In the current work, we are concerned with a type of coupled optical waveguides or cavities alternately carrying gain and loss for the purpose. Attracting wide experimental \cite{5,6,7,8,9,10} and theoretical researches 
\cite{11,12,13,14,15,16,17,18,19,20} recently, these systems governed by non-Hermitian dynamics enjoy the advantage of being capable of changing the light transmission patterns with proper system parameters. Under the balanced gain and loss, these systems manifest a parity-time ($\mathcal{PT}$) symmetry \cite{21}. One property of such systems is that, when they operate in the $\mathcal{PT}$-symmetry broken regime where their coupling intensity is less than the gain/loss rate, the light propagating in the systems can be significantly amplified. Intuitively, the mechanism can be possibly applied to realize entangled output light fields with high intensities, simply by adding an element of small squeezing intensity into one of the coupled components. The interest in the entanglement following non-Hermitian dynamics can also be seen in the study of another system \cite{n4}.

In most of the previous studies, the light fields in $\mathcal{PT}$-symmetric systems are treated as classical electromagnetic
fields. When dealing with the entanglement of the light fields, one will also encounter an indispensable factor accompanying their amplification and damping---the quantum noises acting as the random drives from the associated reservoirs.
The quantum noise operators preserve the proper commutation relations of the evolved light field opertators. 
So far only few recent studies \cite{22, n1, n2} have considered the effects of the quantum noises in optical $\mathcal{PT}$-symmetric systems, including the hybrid ones with other physical elements added into the systems \cite{23, n3}. As it is well known, quantum entanglement is fragile under the influence of the noises from environment \cite{24,25}. How they affect the entanglement generated by $\mathcal{PT}$-symmetric systems was still an open question. In this paper, we will present a study of the problem by quantitatively examining the influence of the quantum noises on the generated entanglement. The comparison between the generated macroscopic quantum entanglement in the absence and presence of amplification/dissipation quantum noise enables one to understand their effects in $\mathcal{PT}$-symmetric systems more deeply.

The rest of the paper is organized as follows. In Sec.\,II, we give a rather detailed account of the dynamical process to create the macroscopic entanglement with a hybrid $\mathcal{PT}$-symmetric system. The solution to the dynamical equation is presented in terms of the quadratures of the output fields. To illustrate the effect of the added squeezing element, we first discuss how it will influence the output photon numbers in Sec.\,III. The interplay between the squeezing and quantum noises is also illustrated with output photon number evolutions. Our main results about the CV entanglement are presented in Sec.\,IV, where the entanglement generated with three different configurations of the system is illustrated under various parametric choices for the system. The different degrees of entanglement generated in the absence and in the presence of the quantum noises are compared. Finally, a brief conclusion is made in Sec.\,V.

\section{Dynamics of a hybrid $\mathcal{PT}$-symmetric system}

\begin{figure}[b!]
\includegraphics[width=\linewidth]{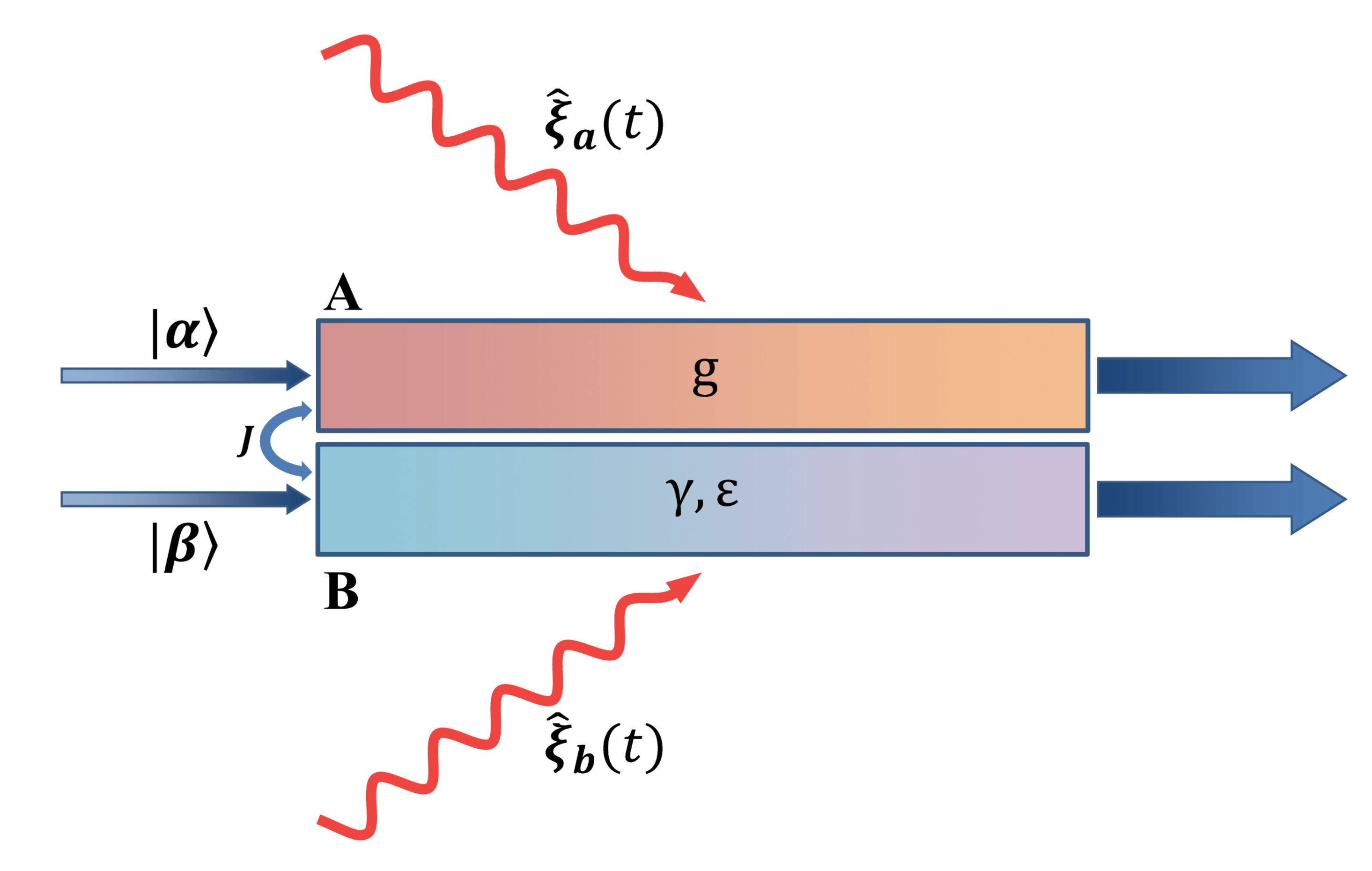}
\caption{The coupled gain-loss waveguide system with an added squeezing element into one of the waveguides. 
The input light fields are in coherent states. Here the squeezing element is supposed to be in the loss channel, without showing the used pumping fields for the amplifications.}
\label{Fig. 1}
\end{figure}

\noindent Due to the inevitable interaction with their environment, most physical systems are open, 
and hence their dynamics becomes non-Hermitian. The system of two coupled waveguides carrying optical gain and loss respectively as in Fig.\,1 is an example of such open system interacting with the bosonic baths. The waveguide $A (B)$ carries a gain (loss) medium, in which the single-mode field $\hat{a}(\hat{b})$ propagates. Here we only consider the propagation of the light fields at a normal group velocity $v_g$, neglecting the possible superluminal propagation of the evanescent wave as described in, e.g., \cite{super01, super02, super1}. The two light fields couple between the waveguides and the effective coupling intensity $J$ can be adjusted by the gap distance. 
The magnitudes of the light field gain rate $g$ and loss rate $\gamma$ are decided by the used media, and the optical gain can be realized by doping Erbium ions into the material \cite{dope}. Moreover, we neglect the gain saturation by assuming a high saturation intensity for the used gain medium (the inclusion of the realistic gain saturation will not change the qualitative picture for our concerned processes). When they are balanced ($g=-\gamma$), there will be a $\mathcal{PT}$ symmetry for the system, as manifested by the effective non-Hermitian Hamiltonian
\begin{equation}
H_\text{PT}=ig\hat{a}^\dagger \hat{a}-ig\hat{b}^\dagger \hat{b}+J\,\big(\hat{a}^\dagger \hat{b}+\hat{a} \hat{b}^\dagger\big).
\label{pt}
\end{equation}
This Hamiltonian is invariant under the simultaneous parity transformation $\hat{a}\leftrightarrow \hat{b}$ and time inversion transformation $i\leftrightarrow -i$, hence the term $\mathcal{PT}$ symmetry. The Hamiltonian comes from the dynamical equations in the previous studies of classical $\mathcal{PT}$-symmetric systems, where the system modes $\hat{a}$ and $\hat{b}$ are replaced by the corresponding classical fields. When $g<J$ ($\mathcal{PT}$-symmetric regime), this Hamiltonian's eigenvalues are imaginary and the transmitting light intensities (proportional to $\langle \hat{a}^\dagger\hat{a}(t)\rangle$ and $\langle \hat{b}^\dagger\hat{b}(t)\rangle$) in both waveguides demonstrate periodic oscillations in time. When $g>J$ ($\mathcal{PT}$ symmetry broken regime), the eigenvalues become real, and the intensities of the transmitting light fields change from oscillatory to exponentially growing. The transition takes place at $g=J$, the exceptional point. 

One problem in applying the above Hamiltonian to deal with the light fields' quantum properties in the concerned systems is that it does not possess the quantum noise elements, which are important to their entanglement.
The amplification and dissipation of the light fields are accompanied by the actions of their associated quantum noises $\hat{\xi}_a(t)$ and $\hat{\xi}_b(t)$; see Fig.\,1. To include their effects we will adopt the following stochastic Hamiltonian 
\begin{equation}
\begin{split}
H_\text{I}=J\big(\hat{a}^\dagger \hat{b}+\hat{a} \hat{b}^\dagger\big)
+ i\sqrt{2g}\;\Big[\hat{a}^\dagger\hat{\xi}^\dagger_a(t)-\hat{a}\, \hat{\xi_a}(t)\Big]\\
+i\sqrt{2\gamma}\;\Big[\hat{b}^\dagger \hat{\xi_b}(t)-\hat{b}\, \hat{\xi}^\dagger_b(t)\Big].
\end{split}
\label{hh}
\end{equation}
The detailed derivation of this stochastic Hamiltonian is given in \cite{23}, and a different notation for the amplification part is used in \cite{22}. 
The operators of these quantum noises satisfy the following relations:
\begin{equation}
\begin{split}
\langle \hat{\xi}^\dagger_c(t) \hat{\xi}_c(t') \rangle=0,\\ 
\langle \hat{\xi}_c(t) \hat{\xi}^\dagger_c(t') \rangle=\delta(t-t'),\\
[\hat{\xi}_c(t), \hat{\xi}^\dagger_c(t')]=\delta(t-t'),
\end{split}
\end{equation}
where $c=a, b$. 
Compared with those derived with Eq. (\ref{pt}), the dynamical equations from the Hamiltonian in Eq. (\ref{hh}) contain the extra quantum noise drive terms. 

To entangle the light fields, one could add a squeezing element into the concerned waveguide system. The action of the squeezing element with the parameter $\epsilon=r\exp(i \theta)$ is described by the Hamiltonian (when it is added into waveguide $B$) 
\begin{equation}
\begin{split}
H_\text{S}=\frac{i}{2}\Big[ \epsilon(\hat{b}^\dagger)^2 -{\epsilon^\ast} (\hat{b})^2 \Big],
\end{split}
\end{equation}
which is from the undeplete pump approximation for a process of second harmonic generation in nonlinear crystal with certain symmetry (for example, LiNbO$_3$) \cite{ca}. Similar squeezing element was also proposed for engineering the quantum properties of other systems (see, e.g. \cite{h-a}).  

All properties of the light fields propagating in the concerned system are the results of their dynamical evolution due to the Hamiltonian $H=H_\text{I}+H_\text{S}$.
To solve their dynamical equations more efficiently, it is convenient to work with the quadratures of the light fields and quantum noises defined as
\begin{equation}
\begin{split}
\hat{q}_c=\frac{1}{\sqrt{2}}\big(\hat{c}+\hat{c}^\dagger\big),\\
\hat{p}_c=-\frac{i}{\sqrt{2}}\big(\hat{c}-\hat{c}^\dagger\big),\\
\hat{Q}_c=\frac{1}{\sqrt{2}}\Big(\hat{\xi_c}(t)+\hat{\xi}^\dagger_c(t)\Big),\\
\hat{P}_c=-\frac{i}{\sqrt{2}}\Big(\hat{\xi_c}(t)-\hat{\xi}^\dagger_c(t)\Big),
\end{split}
\end{equation}
where $c=a, b$. Then the Heisenberg-Langevin equation for the process in Fig.\,1 reads \cite{28}
\begin{equation}
\frac{\mathrm{d}}{\mathrm{d}t}
{\hat{\mathrm{\textbf X}}}(t)=M{\hat{\mathrm{\textbf X}}}(t)
+{\hat{\mathrm{\textbf F}}}(t)
\label{dynamic}
\end{equation}
where
\begin{equation}
{\hat{\mathrm{\textbf X}}}(t)=
\Big( \hat{q}_{a}(t),\, \hat{p}_{a}(t),\,\hat{q}_{b}(t),\, \hat{p}_{b}(t) \Big)^T,
\end{equation}
\begin{equation}
M=\begin{pmatrix}
g & 0 & 0 & J \\
0 & g & -J & 0 \\
0 & J & -\gamma+r\cos \theta & r\sin \theta \\
-J & 0 & r\sin \theta & -\gamma-r\cos \theta
\end{pmatrix}
\label{m1}
\end{equation}
is the dynamic matrix, and
\begin{equation}
\begin{split}
{\hat{\mathrm{\textbf F}}}(t)=\Big(
\sqrt {2g}\;\hat{Q}_{a}(t),
-\sqrt {2g}\;\hat{P}_{a}(t),
\sqrt {2\gamma}\;\hat{Q}_{b}(t),
\sqrt {2\gamma}\;\hat{P}_{b}(t) 
\Big)^T.
\end{split}
\end{equation}
The solution to the dynamical equations takes the form
\begin{equation}
\hat{\mathrm{\textbf X}}(t)=\mathrm {e}^{M t}{\hat{\mathrm{\textbf X}}(0)}+\int_{0}^{t}\mathrm {e}^{M (t-t')}{\hat{\mathrm{\textbf F}}}(t')\; \mathrm{d}t'.
\label{sol}
\end{equation}
The input fields are coherent states, and the evolved quantum states according to the above dynamical equation will be preserved to be Gaussian states. All properties of such Gaussian states can be depicted with the covariance matrix (CM) \cite{3, 4, ade}:
\begin{equation}
V=\begin{pmatrix}
A & C \\
C^T & B \\
\end{pmatrix}
\label{cm}
\end{equation}
where
\begin{equation}
V_{ij}=\langle \hat{X}_i \hat{X}_j+ \hat{X}_j \hat{X}_i\rangle-2\langle\hat{X}_i \rangle\langle\hat{X}_j \rangle.
\label{cm1}
\end{equation}
The expectation values of the homogeneous part in (\ref{sol}) are calculated with respect to the input coherent states $|\alpha,\beta\rangle$, while those of the inhomogeneous part are found with respect to the total reservoir state $\rho_\text{R}=|0\rangle \langle 0|$ (a zero temperature for the reservoirs is assumed). These CM elements can also 
be experimentally measured \cite{detect2}.

\begin{figure}[b!]
	\includegraphics[width=\linewidth]{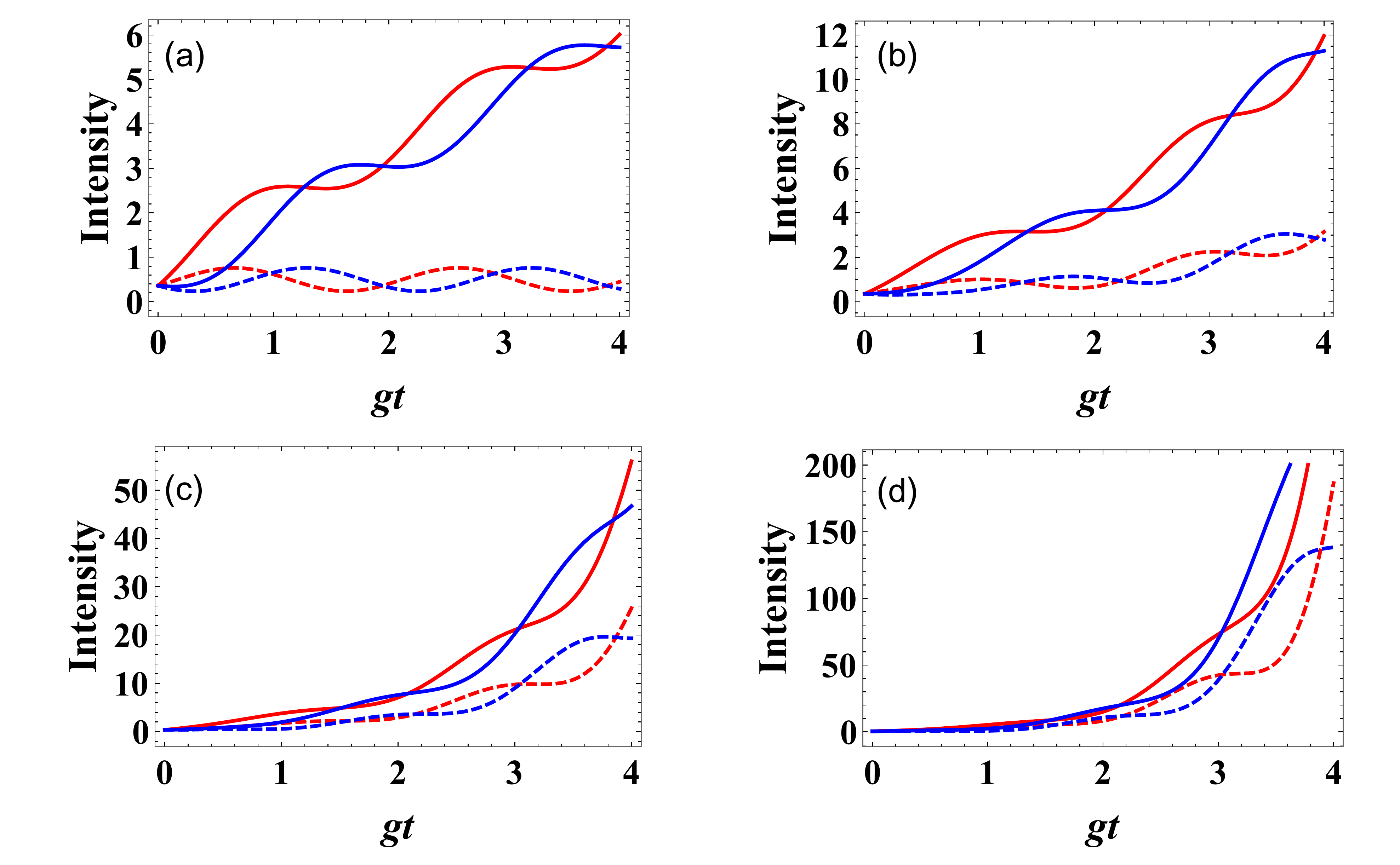}
	\caption{The light intensity $I_a$ and $I_b$ (proportional to the respective photon numbers) out of the gain and loss waveguide, represented by red and blue curves, respectively, as compared with the corresponding quantities $I_{a,h}$, and $I_{b,h}$ calculated without the noise drives (the dashed red and blue curves). Here the dimensionless time $gt$ is used to indicate how long the light fields evolve in the waveguides. We set the parameters as $\theta=\pi/4$, $J=1.9g$, and $g=-\gamma$. The squeezing parameter: (a) $r=0$, (b) $r=0.5g$, (c) $r=g$, and (d) $r=1.5g$. The input coherent states are given as $\alpha=\beta=0.6\exp(i\pi/4)$. }
	\label{Fig. 2}
\end{figure}
%%%%%%%%%%%%%%%%%%%%%%%%%%%%%
\section{Evolution of photon number and waveguide-mode correlation}
\noindent  
A main purpose of the current study is to find out how the quantum noises will affect the entanglement generated with the setup in Fig.\,1. To see this, one could compare the values of the entanglement found as the results of the evolutions according to the Hamiltonian in Eq. (\ref{pt}) and in Eq. (\ref{hh}), respectively. As we have mentioned in the above, the only difference in the Heisenberg-Langevin equation derived with the latter is an extra quantum noise drive 
term, ${\hat{\mathrm{\textbf F}}}(t)$ in Eq. (\ref{dynamic}), which consists of the components of pure random variables.
By intuition, such random drives from the environment could only modify the dynamics of the system without changing the evolution patterns of the measurable quantities so much, as it has been found from the photon number evolutions in 
a $\mathcal{PT}$-symmetric system without squeezing \cite{22}.      

A relevant question is whether the added squeezing will make a considerable difference. 
To answer the question, we examine how the output light fields' intensities evolve according to the full dynamical equation, Eq. (\ref{dynamic}). Previously, the evolved light intensities in a $\mathcal{PT}$-symmetric system without squeezing have been studied for input single photon and vacuum state \cite{22}. Due to the amplification noise, the output field intensity is found not to be zero even when the input field is in a vacuum state. After adding a squeezing element, 
we find that the photon numbers can be enhanced further, in addition to the effect of the gain medium at the rate $g$. 
Fig.\,2 shows the intensities plotted for a setup with the squeezing elements of different intensities in the damping waveguide, indicating that the photon numbers will be intensified by increased squeezing parameter $r$. The contribution 
from the homogeneous part in Eq. (\ref{dynamic}) will become much more significant due to an increasing squeezing parameter $r$, so that the relative difference between the light intensities obtained by using the Hamiltonians in Eq. (\ref{pt}) and in Eq. (\ref{hh}), respectively, can become smaller than that in the previously studied situation without squeezing; compare the results in Figs.\,2(b)-2(d) with those in Fig.\,2(a). Here we only show the results in the regime $J>g$ because those for $J<g$ are similar. Similar to the situation of a simple $\mathcal{PT}$-symmetric system without squeezing, the existence of the quantum noises simply quantitatively modifies an output light intensity. It is an intriguing issue whether the nonlocal properties like entanglement also behave similarly under the quantum noises. 

We continue to look at a nonlocal quantity as the correlation function defined as 
$|\langle \hat{a}^\dagger \hat{b}\rangle-\langle \hat{a}^\dagger\rangle \langle \hat{b} \rangle|$, assuming that the squeezing element is inside the damping waveguide.
The amplification noise can significantly contribute to this function.
On the surface, such correlation function, which evolves with time, is more or less relevant to the entanglement of the light fields. Fig.\,3 illustrates the evolution of the function for three different squeezing parameters. As one should expect, 
the correlation becomes stronger with a larger squeezing parameter $r$. Does this phenomenon reflects a similar pattern for the corresponding entanglement between the output light fields? 

\begin{figure}
	\includegraphics[width=\linewidth]{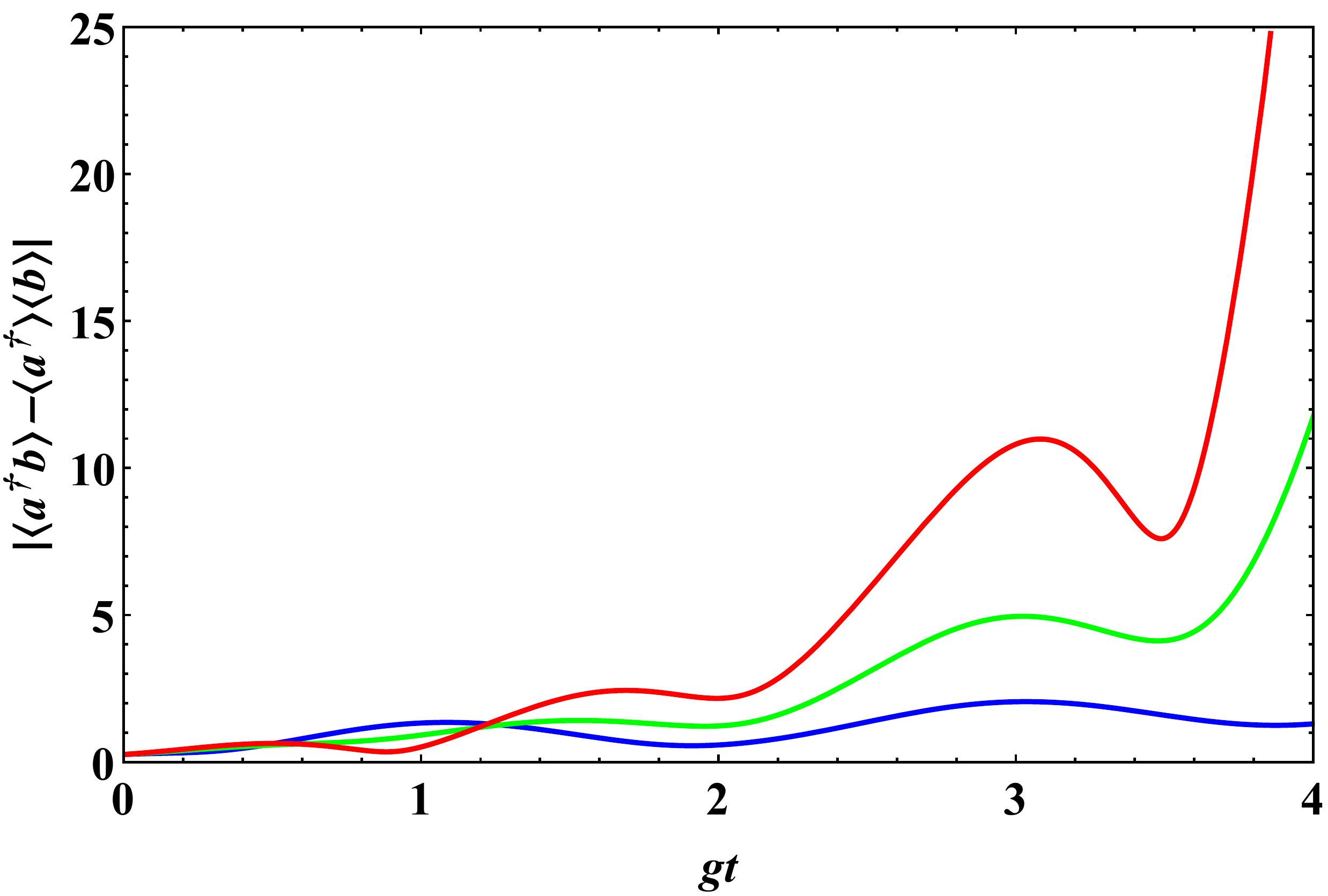}
	\caption{The correlation function $|\langle \hat{a}^\dagger \hat{b}\rangle-\langle \hat{a}^\dagger\rangle \langle \hat{b} \rangle|$ calculated with $J=1.9g$. We set $g=-\gamma$ and $\theta=\pi/4$. Three different values of squeezing, $r=0$ (blue), $r=g$ (green), and $r=1.5g$ (red), are considered for the squeezing element in the damping waveguide. The input coherent states are given as $\alpha=\beta=0.6\exp(i\pi/4)$.}
	\label{Fig. 3}
\end{figure}
%%%%%%%%%%%%%%%%%%%%%%%%%%%%%%
\section{entanglement of output fields}
\noindent  
If the entanglement of two output light fields can also grow with time, as the correlation function in Fig.\,3, 
it will be possible to realize highly entangled strong light fields with ease.
The degrees of the possibly generated entanglement of Gaussian states can be measured by the logarithmic negativity calculated with the CM of the the two system modes:
\begin{equation}
E_N=\mathrm{max}[0, -\ln \eta]
\end{equation}
where
\begin{equation}
\eta=\frac{1}{\sqrt 2}\sqrt{\Sigma-\sqrt{\Sigma^2-4\,\text{det}\,V}}
\end{equation}
and
\begin{equation}
\Sigma=\text{det}\,A+\text{det}\,B-2\,\text{det}\,C.
\end{equation}
The notations for the CM defined in Eq.\,(\ref{cm}) are used here. 
There is an important property of such entanglement for the input light fields in coherent states.
In the calculation of the elements of CM, the contribution from the homogeneous part in Eq.\,(\ref{sol}) only manifests from the commutation of the mode operators, after deducting the mean values $\langle\hat{X}_i \rangle\langle\hat{X}_j \rangle$ in Eq.\,(\ref{cm1}). As a result, the intensity of the input coherent states will become irrelevant to the entanglement, the degree of which is mainly influenced by the noise drives leading to the inhomogeneous part in Eq.\,(\ref{sol}). 

In what follows, we will study the output entanglement from three distinct setups---the squeezing element is in the damping waveguide, in the amplification waveguide, and in both waveguides.

\subsection{Squeezing element in the damping waveguide}
\noindent The first situation is to add the squeezing element into the damping waveguide as in 
Fig.\,1. Without considering the noise drive term ${\bf \hat F}(t)$ that gives the inhomogeneous part in Eq.\,(\ref{sol}), the output fields will become strongly entangled; see Fig.\,4. In the $\cal{PT}$-symmetric regime as illustrated in Fig.\,4(a), the entanglement values are oscillating; in the symmetry broken regime as in Fig.\,4(b), the entanglement values continuously grow with time. The calculated entanglement without considering the quantum noises resembles the evolutions of the correlation function in Fig. 3; the larger the squeezing intensity $r$ is, the higher the entanglement will be.

\begin{figure}
	\includegraphics[width=\linewidth]{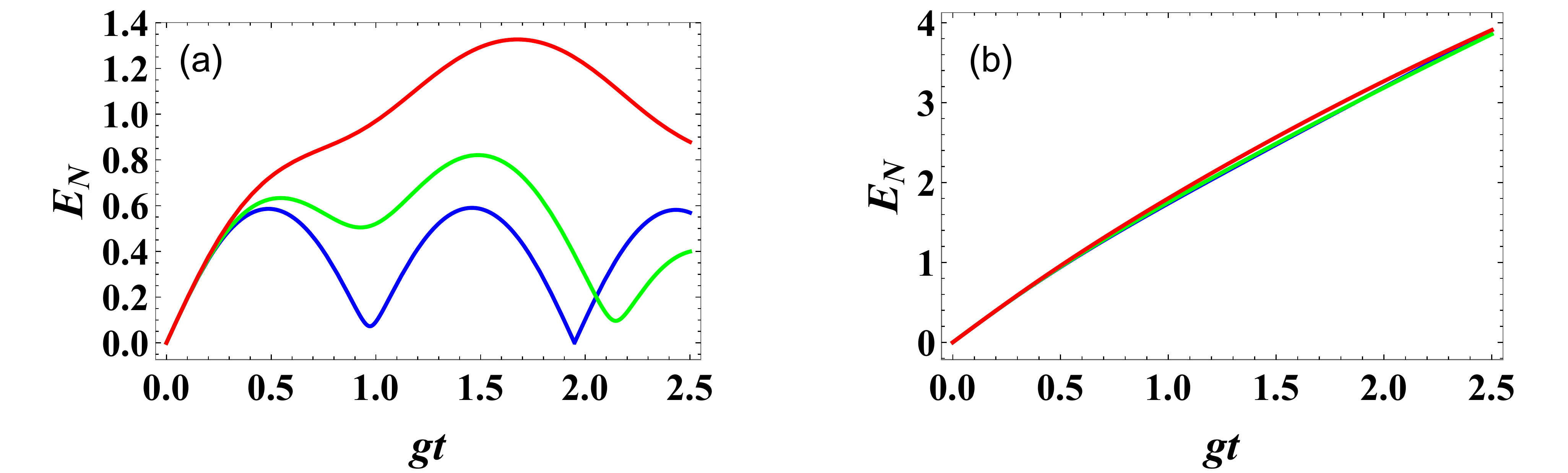}
	\caption{Calculated entanglement values without quantum noise effect, when a squeezing element is put into the damping waveguide. We set $g=-\gamma$ and $\theta=\pi/4$ and $J=1.9g$ in (a), $J=0.7g$ in (b). The squeezing parameters are chosen as $r=0.1g$ (blue), $r=0.7g$ (green), and $r=1.3g$ (red).}
	\label{Fig. 4}
\end{figure}

The realistic evolution of the entanglement, however, radically differs from the pattern in Fig.\,4, after we include the effects of the quantum noises. In the situation of adding a squeezing element into the damping waveguide, the entanglement is found to totally disappear all the time. The action of the quantum noises is simultaneous with the squeezing action entangling the light fields, and dominates over the latter from the beginning due to the interplay between the squeezing and the dissipation noise, which enhances the influence of the dissipation noise on the relevant CM elements [check the dynamic matrix in Eq. (8) and the second term in Eq. (10)]. This result clearly demonstrates the significance of the system's interaction with its environment. Compared with the correlation function illustrated in Fig.\,3, one concludes that the existence of the correlation between the two system modes does not suffice to give rise to their entanglement. In other words, correlation and entanglement for the light fields are not equivalent. The entanglement has to be determined by the relations between all CM elements in Eq.\,(\ref{cm1}) as the similar correlation functions, so its existence should be much more restricted.
%%%%%%%%%%%%%%%%%%%%%%%%%%%%%
\subsection{Squeezing element in the gain waveguide}
\noindent If a squeezing element is inserted into the amplifying waveguide, the dynamic matrix in Eq.\,(8) will be changed to
\begin{equation}
M=\begin{pmatrix}
g+r\cos \theta &  r\sin \theta & 0 & J \\
 r\sin \theta & g-r\cos \theta & -J & 0 \\
0 & J & -\gamma & 0 \\
-J & 0 & 0 & -\gamma 
\end{pmatrix}.
\end{equation}
The dynamical evolutions of the waveguide modes, as given by Eq.\,(\ref{sol}), will be changed accordingly.

Now let us look at how the entanglement between the two waveguide mode will change after the location of the squeezing element is swapped to the gain waveguide. In Fig.\,5 we illustrate the numerically calculated entanglement values in both $\mathcal{PT}$ symmetry broken regime [Figs.\,5(a) and 5(c)] and $\mathcal{PT}$ symmetry preserved regime [Figs.\,5(b) and 5(d)]. One sees that, under the quantum noise effects, there will still be certain amounts of remnant entanglement, though they are much less significant as compared with the predictions without involving the quantum noises. The asymmetry in the interplay of the squeezing effect with the amplification noise and with the dissipation causes the difference. 
The entanglement evolving under the full dynamics also exhibits a phenomenon of entanglement sudden death (ESD) \cite{25}, to have its optimum value or peak value obtained at a certain time (equivalent to an optimum waveguide length).
As a result of the different dynamical evolution from that of putting the squeezing element in the damping waveguide, the entanglement between the waveguide modes exists for a period of time. 

\begin{figure}[h]
	\includegraphics[width=\linewidth]{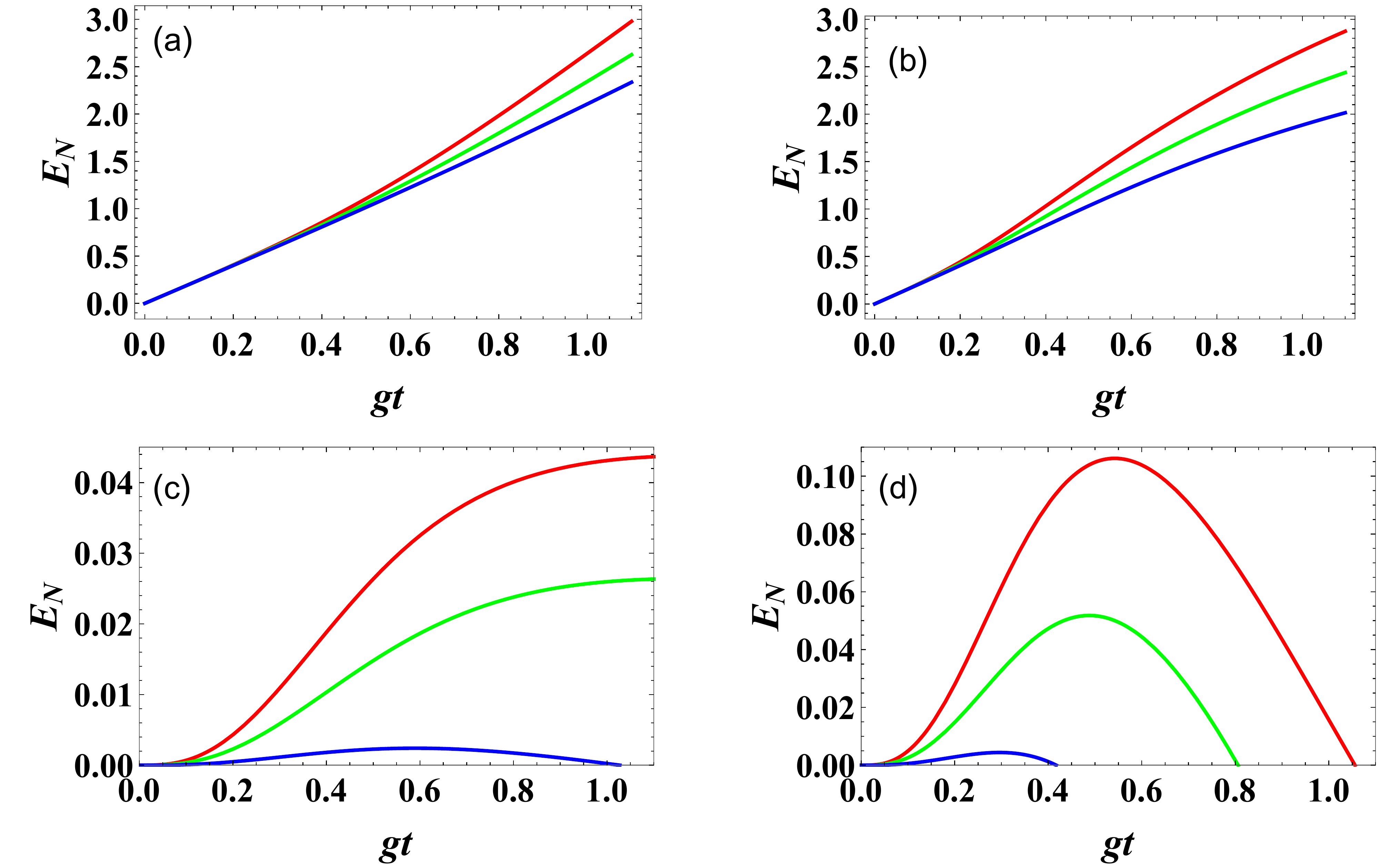}
	\caption{Entanglement generated by a squeezing element in the gain waveguide. Here we choose $\theta=\pi/4$ and $g=-\gamma$. In (a) and (c) 
 $J=0.7g$, and in (b) and (d) $J=1.9g$. The entanglement evolutions in (a) and (b) are found with the non-Hermitian Hamiltonian in Eq. (1), together with the squeezing action. Those in (c) and (d) are found with the full dynamics including the noise actions. The squeezing parameters are $r=3.5g$ (red), $r=3g$ (green), and $r=2.5g$ (blue).}
	\label{Fig. 5}
	\end{figure}

%%%%%%%%%%%%%%%%%%%%%%%%%%%%%%
\subsection{Squeezing elements in both waveguides}

\noindent At last we will consider the situation of adding squeezing elements into both waveguides. 
In this situation, the dynamical matrix for the system is
\begin{equation}
M=\begin{pmatrix}
g+r\cos \theta & r\sin \theta & 0 & J \\
 r\sin \theta & g-r\cos \theta & -J & 0 \\
0 & J & -\gamma+r\cos \theta &  r\sin \theta \\
-J & 0 & r\sin \theta & -\gamma-r\cos \theta
\end{pmatrix}.
\end{equation}
Under the condition $g=-\gamma$, the system exhibits a $\mathcal{PT}$ symmetry like the Hamiltonian in Eq.\,(1).
In this situation, the entanglement will exhibit more different features. For example, under the quantum noise drives, there will be no entanglement 
at all if the system operates in the $\mathcal{PT}$ symmetry broken regime with $J<g$.

In the $\mathcal{PT}$-symmetric regime with $J>g$, there will be entanglement generated with the proper lengths of the waveguides, as demonstrated 
in Fig.\,6. Due to the squeezing elements in both waveguides, the generated entanglement can be considerable as indicated by their values in Fig.\,6(b). 
The entanglement will be created from zero along the propagation of the light fields and will evolve to a point of ESD after a period of time, and hence an optimum amount of entanglement should be obtained with a proper waveguide length.

\vspace{0.5cm}
\begin{figure}[h!]
\includegraphics[width=\linewidth]{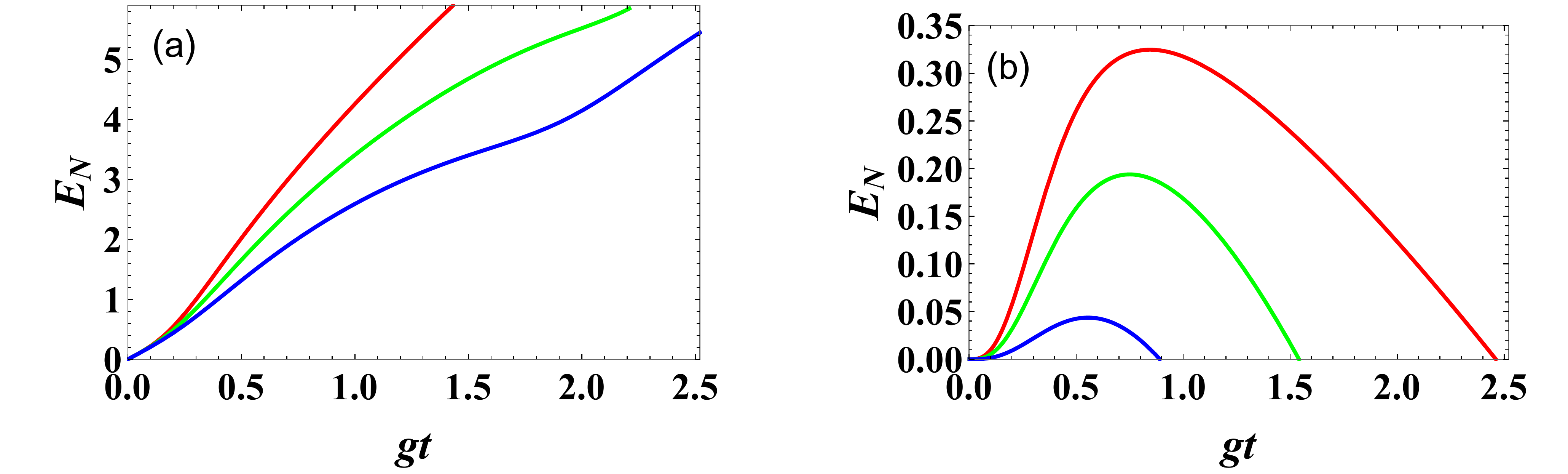}
\caption{Entanglement generated by placing identical squeezing elements into both waveguides. We set $\theta=\pi/4$, $g=-\gamma$ and $J=1.9g$. In (a) the entanglement values are calculated without considering quantum noises, and in (b) the entanglement values are found with the full dynamics. The squeezing parameters are $r=2.7g$ (red), $r=2g$ (green), and $r=1.7g$ (blue).}
\label{Fig. 6}
\end{figure}

\section{Conclusion}
\noindent We have studied the entanglement generated with a hybrid $\mathcal{PT}$-symmetric setup, which is realized by adding a squeezing element into two coupled waveguides respectively carrying the medium with the balanced gain and loss rate. By intuition, the existing quantum noises would slightly modify the dynamics of such system, so that the evolution patterns of the generated entanglement would not be changed so much from those predicted with the non-Hermitian Hamiltonian in Eq.\,(1).
Then highly entangled light fields of significant intensity can be readily created by such setup, especially the system operating in the $\mathcal{PT}$ symmetry broken regime. As a matter of fact, however, the quantum noises can completely kill the entanglement, rendering its evolution totally different from those of photon numbers and field-mode correlations. As indicated by our results found with the full dynamics, a certain amount of the entanglement can be achieved only by placing the squeezing element inside the waveguide amplifying the propagating light field.
The possible experimental realization of such systems relies on finding a material or a method to purely amplify and squeeze the input light at the same time. The importance of studying this model setup is to clarify the fact that quantum noises must be considered in $\mathcal{PT}$-symmetric optical systems for engineering the quantum properties of light fields.   

\begin{acknowledgments}

We acknowledge funding support in part from NSFC (Grant No. 11574093 and Grant No. 61435007). 

\end{acknowledgments}


\begin{thebibliography}{}
\bibitem{1}
M. A. Nielsen and I. L. Chuang, \textit{Quantum Computation and Quantum Information:10th Anniversary Edition}, Cambridge University Press, Cambridge, 2011.
\bibitem{3}
S. L. Braunstein and P. van Loock, Rev.\ Mod.\ Phys. {\bf 77}, 513 (2005). 
\bibitem{4}
C. Weedbrook, S. Pirandola, R. García-Patrón, N. J. Cerf, T. C. Ralph, J. H. Shapiro, and S. Lloyd, Rev.\ Mod.\ Phys. {\bf 84}, 621 (2012)
\bibitem{ade} G. Adesso, A. Serafini, and F. Illuminati, Phys. Rev. A 70(2), 022318 (2004).
\bibitem{ex0} A. I. Lvovsky, R. Ghobadi, A. Chandra, A. S. Prasad, and C. Simon, Nature Physics 9, 541 (2013).
\bibitem{ex1} H. Jeong, A. Zavatta, M. Kang, S.-W. Lee, L. S. Costanzo, S. Grandi, T. C. Ralph, and M. Bellini, Nat. Photonics 8, 564 (2014).
\bibitem{a1} B. He, M. Nadeem, and J. A. Bergou, Phys. Rev. A 79, 035802 (2009).
\bibitem{Raeisi}
S. Raeisi, P. Sekatski, and C. Simon, Phys. Rev. Lett {\bf 107}, 250401 (2011).
\bibitem{a2} A. V. Sharypov and B. He, Phys. Rev. A 87, 032323 (2013).
\bibitem{a3} M. Chekhova, G. Leuchs, and M. Zukowski, Optics Communications 337, 27 (2015).
\bibitem{5}
A. Guo, G. J. Salamo, D. Duchesne, R. Morandotti, M. Volatier-Ravat, V. Aimez, G. A. Siviloglou, and D. N. Christodoulides,
Phys.\ Rev.\ Lett. {\bf 103}, 093902 (2009).
\bibitem{6}
C. E. R\"{u}ter, K.G.Makris, R. El-Ganainy, D.N. Christodoulides,
M. Segev, and D. Kip, Nat. Phys. {\bf 6}, 192 (2010).
\bibitem{7}
A. Regensburger, C. Bersch, M. A. Miri, G. Onishchukov, D. N.
Christodoulides, and U. Peschel, Nature (London) {\bf 488}, 167
(2012).
\bibitem{8}
T. Eichelkraut, R. Heilmann, S.Weimann, S. St\"{u}tzer, F.Dreisow,
D. N. Christodoulides, S. Nolte, and A. Szameit, Nat. Commun.
{\bf 4}, 2533 (2013).
\bibitem{9}
L. Chang, X. Jiang, S. Hua, C. Yang, J.Wen, L. Jiang, G. Li, G.
Wang, and M. Xiao, Nat. Photonics {\bf 8}, 524 (2014).
\bibitem{10}
B. Peng, S. K. O¨ zdemir, F. Lei, F. Monifi, M. Gianfreda, G. L.
Long, S. Fan, F. Nori, C. M. Bender, and L. Yang, Nat. Phys.
{\bf 10}, 394 (2014).
\bibitem{11}
R. El-Ganainy, K. G. Makris, D. N. Christodoulides, and Z. H.
Musslimani, Opt. \ Lett. {\bf 32}, 2632 (2007).
\bibitem{12}
S. Klaiman, U. G\"{u}nther, and N. Moiseyev, Phys.\ Rev.\ Lett. {\bf 101},
080402 (2008).
\bibitem{13}
H. Schomerus, Phys.\ Rev.\ Lett. {\bf 104}, 233601 (2010).
\bibitem{14}
Y. D. Chong, L. Ge, and A. D. Stone, Phys.\ Rev.\ Lett. {\bf 106},
093902 (2011).
\bibitem{15}
Z. Lin, H. Ramezani, T. Eichelkraut, T. Kottos, H. Cao, and D.
N. Christodoulides, Phys.\ Rev.\ Lett. {\bf 106}, 213901 (2011).
\bibitem{16}
C. M. Bender, M. Gianfreda, S. K. O\''{z}demir, B. Peng, and L.
Yang, Phys.\ Rev. A {\bf 88}, 062111 (2013).
\bibitem{17}
X. Luo, J. Huang, H. Zhong, X. Qin, Q. Xie, Yu. S. Kivshar, and
C. Lee, Phys.\ Rev.\ Lett. {\bf 110}, 243902 (2013).
\bibitem{18}
R. El-Ganainy, M. Khajavikhan, and L. Ge, Phys.\ Rev. A {\bf 90},
013802 (2014).
\bibitem{19}
M. H. Teimourpour, R. El-Ganainy, A. Eisfeld, A. Szameit, and
D. N. Christodoulides, Phys.\ Rev. A {\bf 90}, 053817 (2014).
\bibitem{20}
S. Longhi, Opt. Lett. {\bf 40}, 5694 (2015).
\bibitem{21}
C. M. Bender and S. Boettcher, Phys.\ Rev.\ Lett. {\bf 80}, 5243 (1998).
\bibitem{n4}T. E. Lee, F. Reiter, and N. Moiseyev, Phys. Rev. Lett. 113, 250401 (2014).
\bibitem{28}
C. W. Gardiner and P. Zoller, \textit{Quantum Noise}, Springer-Verlag,
Berlin, 2010.
\bibitem{22}
G. S. Agarwal and K. Qu, Phys.\ Rev. A {\bf 85}, 031802(R) (2012).
\bibitem{n1} B. He, L. Yang, Z. Zhang, and M. Xiao, Phys. Rev. A 91, 033830 (2015).
\bibitem{n2} K. V. Kepesidis, T. J. Milburn, J. Huber, K. G. Makris, S. Rotter,
and P. Rabl, New J. Phys. {\bf 18}, 095003 (2016).
\bibitem{23}
B. He, S.-B. Yan, J. Wang, and M. Xiao, Phys.\ Rev. A {\bf 91}, 053832 (2015).
\bibitem{n3} B. He, L. Yang, and M. Xiao, Phys. Rev. A {\bf 94}, 031802(R) (2016).
\bibitem{24}
T. Yu and J. H. Eberly, Phys.\ Rev.\ Lett. {\bf 93}, 140404 (2004).
\bibitem{25}
T. Yu and J. H. Eberly, Science, {\bf 323}, 598 (2009).
\bibitem{super01} A. A. Stahlhofen and G. Nimtz, Europhys. Lett. 76, 189
(2006).
\bibitem{super02} G. Nimtz, Lect. Notes Phys. {\bf 702}, 509 (2006).
\bibitem{super1} Z.-Y. Wang, C.-D. Xiong, and B. He, Phys. Rev. A {\bf 75}, 013813 (2007).
\bibitem{dope} P. C. Becker, N. A.  Olsson, and J. R. Simpson, \textit{Erbium-doped Amplifiers Fundamentals and Technology}, Academic press, 1999.
\bibitem{ca} H. J. Carmichael, \textit{Statistical Methods in Quantum Optics 2}, Springer-Verlag, 2008.
\bibitem{h-a} G. S. Agarwal and S. Huang, Phys. Rev. A {\bf 93}, 043844 (2016).
\bibitem{detect2}V. D'Auria, S. Fornaro, A. Porzio, S. Solimeno, S. Olivares, and M. G. A. Paris, Phys. Rev. Lett. 102, 020502 (2009).
\end{thebibliography}
\end{document}